\documentclass{PoS}

\title{The chiral condensate from the Dirac spectrum in BSM gauge theories}

\ShortTitle{BSM chiral condensate from Dirac spectrum}

\author{Zolt\'an Fodor\\
        Department of Physics, University of Wuppertal,
        Gau$\beta$strasse 20, D-42119, Germany\\
        J\"ulich Supercomputing Center, Forschungszentrum,
        J\"ulich, D-52425 J\"ulich, Germany\\
        Email: \email{fodor@bodri.elte.hu}}

\author{\speaker{Kieran Holland}\\
       Albert Einstein Center for Fundamental Physics, Institute for
        Theoretical Physics, \\
        Bern University, Sidlerstrasse 5, CH-3012 Bern, Switzerland\\
      Department of Physics, University of the Pacific,
        3601 Pacific Ave, Stockton CA 95211, USA\\
        Email: \email{kholland@pacific.edu}}

\author{Julius Kuti\\
       Department of Physics 0319, University of California, San Diego\\
        9500 Gilman Drive, La Jolla, CA 92093, USA\\
        E-mail: \email{jkuti@ucsd.edu}}

\author{D\'aniel N\'ogr\'adi\\
        Institute for Theoretical Physics, E\"otv\"os University,
        H-1117 Budapest, Hungary\\
        Email: \email{nogradi@bodri.elte.hu}}
        
\author{Chik Him Wong\\
        Department of Physics 0319, University of California, San Diego\\
        9500 Gilman Drive, La Jolla, CA 92093, USA\\
        E-mail: \email{rickywong@physics.ucsd.edu} }

\abstract{The eigenvalues of the Dirac operator at finite volume encode whether or not chiral symmetry is spontaneously broken in a massless theory. We apply this framework in a particular BSM context, namely $SU(3)$ gauge theory with $N_f=2$ massless flavors in the 2-index symmetric (sextet) representation. Our first results are at a single lattice spacing. We find that both the density of near-zero eigenvalues and the renormalization group invariant mode number indicate spontaneous symmetry breaking. Quantitatively, there is a discrepancy between the determination of the fermion condensate in the chiral limit via the eigenvalue spectrum and the determinations from direct measurements of the chiral condensate and the GMOR relation. We comment on possible explanations of this discrepancy and further refinements of this study.}

\FullConference{31st International Symposium on Lattice Field Theory - LATTICE 2013\\
		July 29 - August 3, 2013\\
		Mainz, Germany}

\begin{document}

\section{Chiral symmetry breaking}

One possible dynamical explanation for the newly-observed 125~GeV boson is a composite Higgs model driven by an underlying strongly-interacting gauge theory~\cite{Weinberg:1979bn}. Spontaneously broken chiral symmetry in the gauge theory manifests itself via Goldstone bosons which become the longitudinal modes of the $W^{\pm}$ and $Z$ gauge bosons. The composite scalar, a Higgs impostor, could be a light and narrow state if the underlying gauge theory is near-conformal. The minimal realization of the composite Higgs scenario is $SU(3)$ gauge theory with $N_f=2$ massless flavors in the 2-index symmetric (sextet) representation~\cite{Sannino:2004qp}. Perturbation theory and various approximation techniques suggest this model is close to, or possibly in, the region of conformality. Our non-perturbative lattice study of the mass spectrum is consistent with near-conformality and spontaneous symmetry breaking~\cite{Fodor:2012ty}, in agreement with finite-temperature studies~\cite{Kogut:2011ty} and also with a small $\beta$ function of the renormalized gauge coupling which is possibly nonzero~\cite{DeGrand:2010na}. Of particular interest is our recent work with evidence of a light composite scalar in this model~\cite {Ricky}. The gauge theory with this novel fermion representation is an economic BSM theory with no extraneous light degrees of freedom. 

\begin{figure}[thb!]
\begin{center}
\begin{tabular}{cc}
\includegraphics[width=0.49\textwidth]{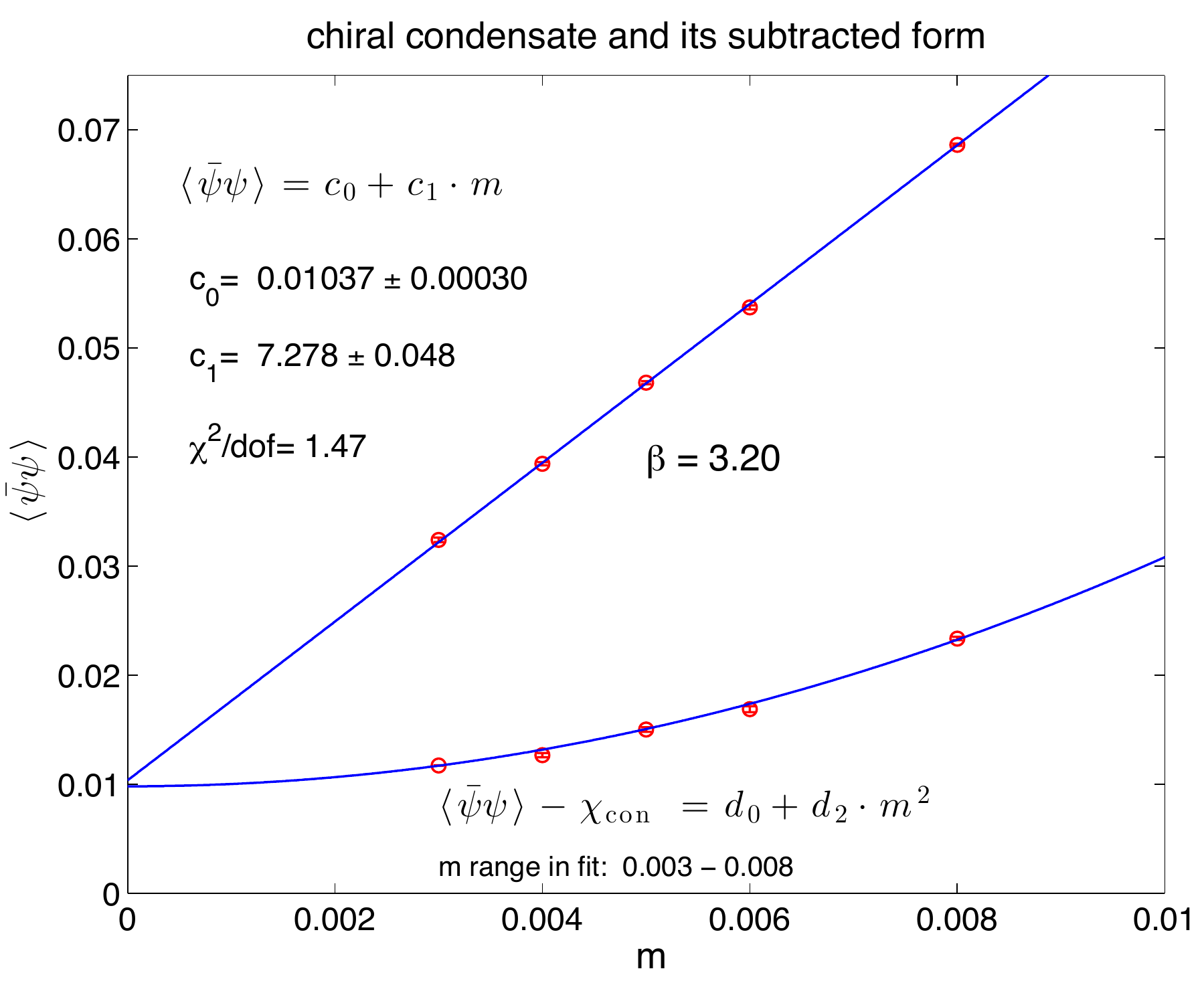}&
\includegraphics[width=0.47\textwidth]{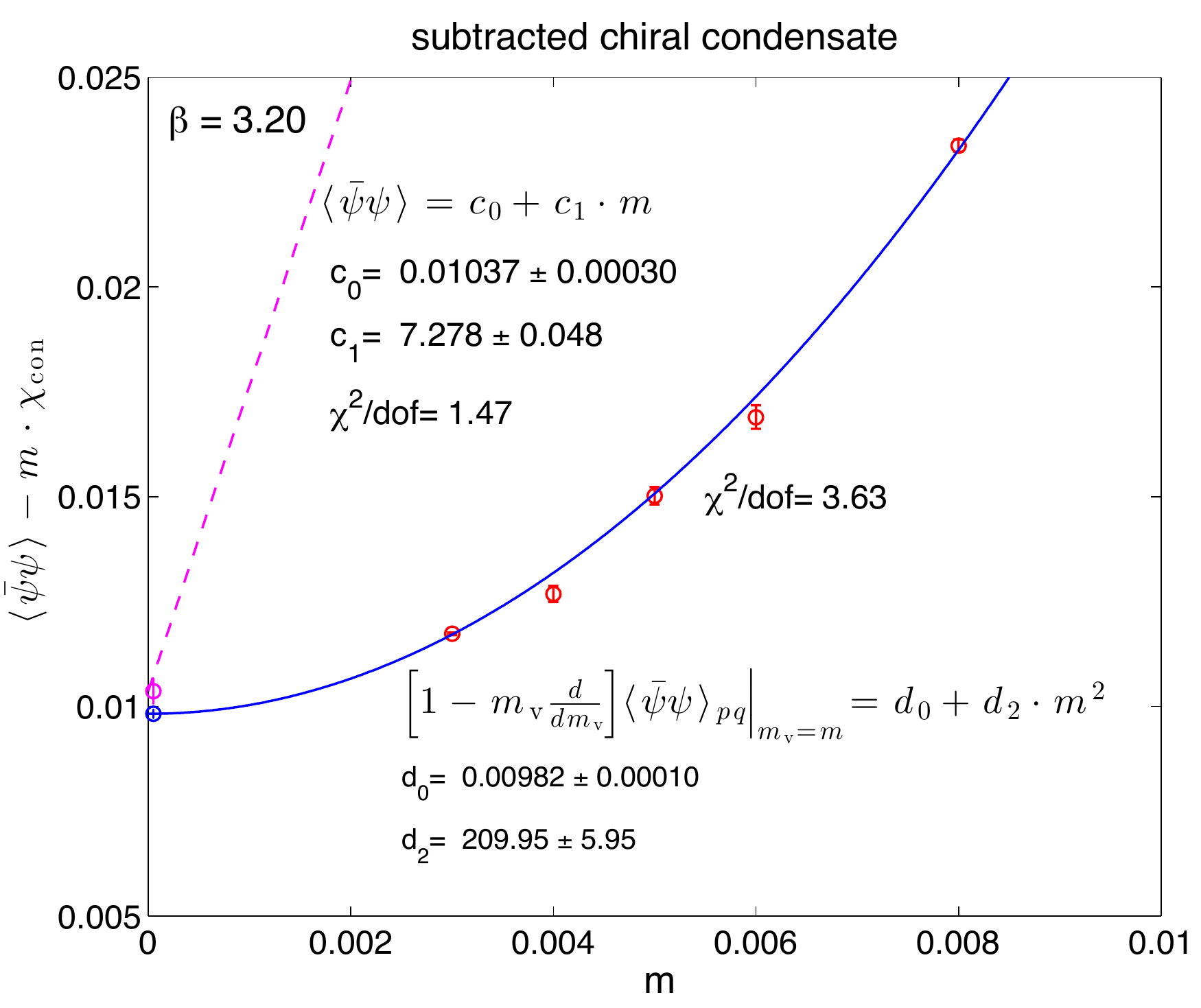}
\end{tabular}
\end{center}
\vskip -0.3in
\caption{\footnotesize (left) Chiral extrapolation of the fermion condensate and its reduced form with a subtracted derivative. The condensate data are well described by a linear fit and higher order terms cannot be detected with significant accuracy. (right) Magnification of the reduced condensate data. The extrapolation excludes a linear term, which is approximately removed by the derivative subtraction, reducing the steep mass dependence. As required, the condensate and reduced condensate give consistent values in the chiral limit.}
\label{fig0}
\end{figure}

A natural way to decide if chiral symmetry is spontaneously broken is to measure via lattice simulations the fermion condensate $\langle \bar{\psi} \psi \rangle$ at finite fermion mass $m$ and find its limiting behavior towards the chiral limit. The practical difficulty with this method is the steep mass dependence of the condensate due to UV-divergent contributions. Even with small statistical errors, the extrapolation to zero mass is challenging. We have in addition used a separate independent condensate observable from which the leading UV-divergent contribution is significantly suppressed, reducing the mass dependence. This operator and the unmodified condensate extrapolate to the same chiral limit as shown in Figure~\ref{fig0}. From our simulations we find that both observables consistently indicate a non-zero fermion condensate in the chiral limit and spontaneous symmetry breaking, we wish to shore up this conclusion using other methods.

One consistency check is the GMOR relation, which connects the fermion condensate to the pseudo-Goldstone mass $M_\pi$ and decay constant $F_\pi$ as $\langle \bar{\psi} \psi \rangle= M_\pi^2 F_\pi^2/m$ for two flavors summed. We show the results of our mass spectrum analysis on the left in Figure~\ref{fig1}. The variation with mass is less than for the directly-measured condensate, but is still sizable. A quadratic extrapolation to the chiral limit describes the data well, which is consistent with independent fits of $M_\pi^2$ and $F_\pi$. However the extrapolated value from the GMOR relation differs from the extrapolations of the directly-measured original and subtracted condensate. One possible explanation of the discrepancy is that taste breaking may invalidate using the continuum GMOR relation, which could be remedied by implementing staggered chiral perturbation theory or by running additional simulations at smaller lattice spacing. The chiral theory might also need to be enlarged to include the effects of the low-lying scalar state.  

\section{Eigenvalues}

\begin{figure}[thb!]
\begin{center}
\begin{tabular}{cc}
\includegraphics[width=0.46\textwidth]{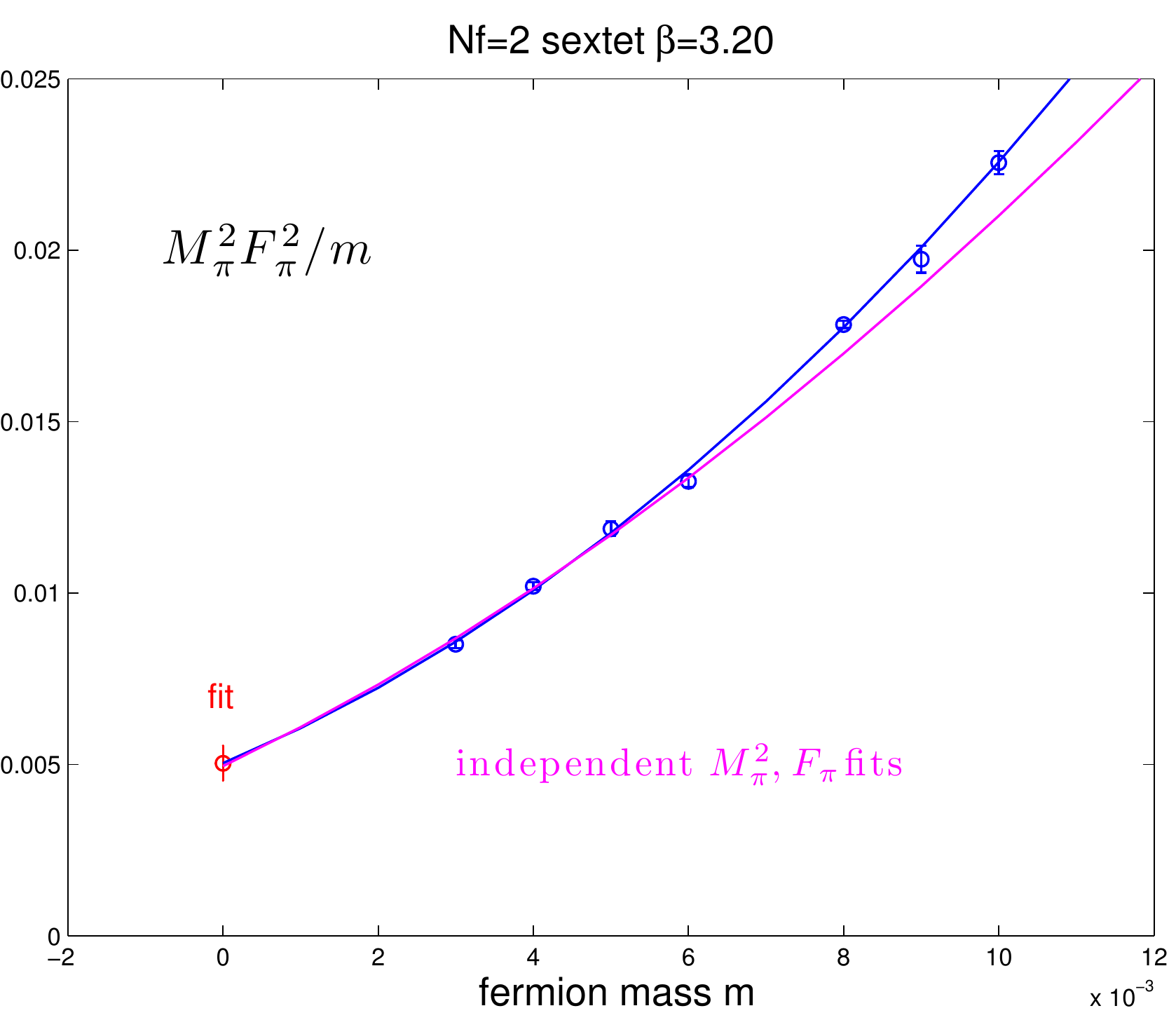}&
\includegraphics[width=0.49\textwidth]{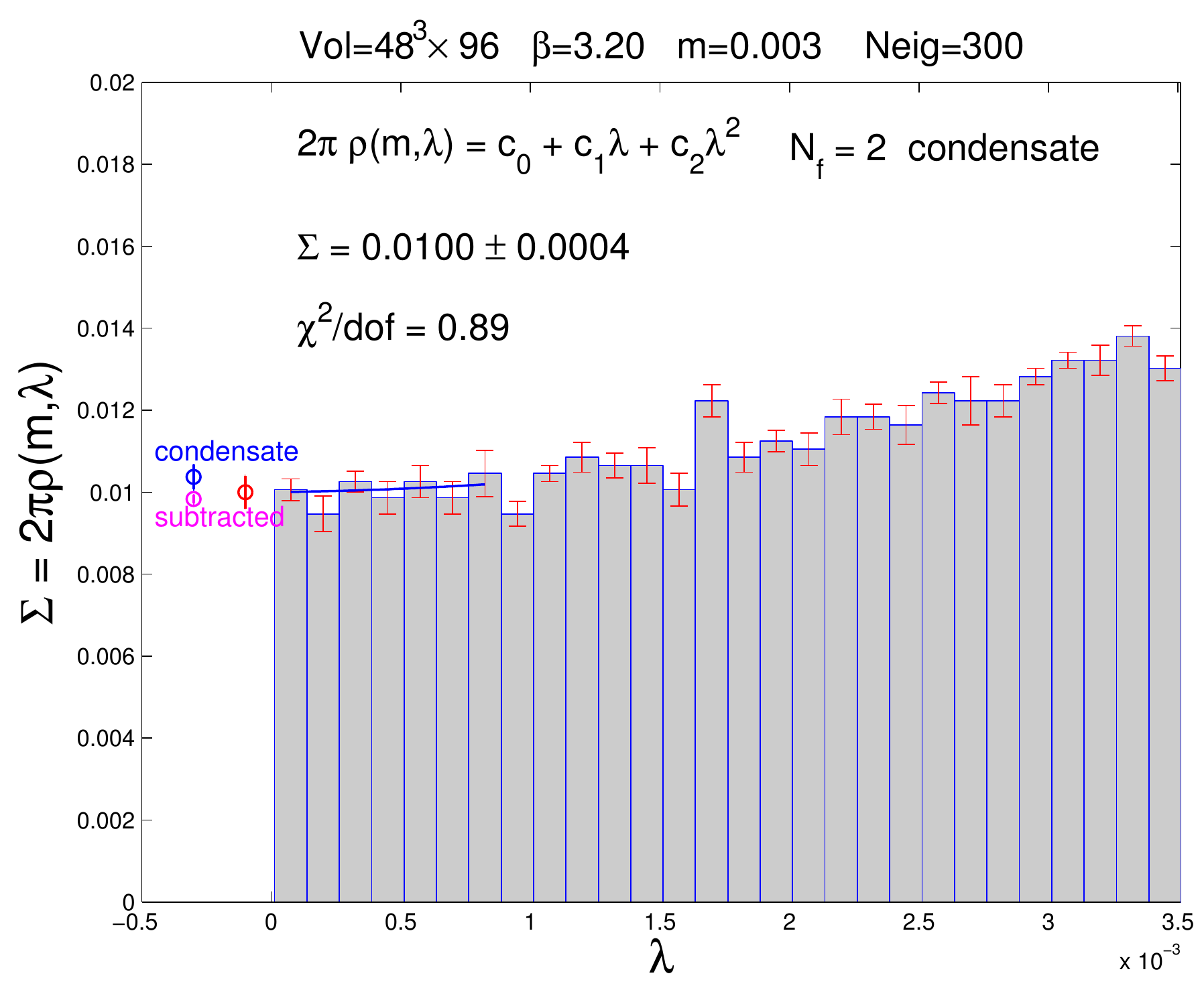}
\end{tabular}
\end{center}
\vskip -0.3in
\caption{\footnotesize (left) The fermion condensate as determined via the GMOR relation. The blue curve is a quadratic fit of all ratio data for $m = 0.003$ -- 0.008, the magenta curve is a combination of separate fits of $M_\pi^2$ and $F_\pi$ for $m = 0.003$ -- 0.006. (right) The eigenvalue density measured on $48^3 \times 96$ lattice volumes. The extrapolation of the eigenvalue density (red) is compared with the directly measured condensate $\langle \bar{\psi} \psi \rangle$ (blue) and the subtracted condensate (magenta) from Figure~1.}
\label{fig1}
\end{figure}

The eigenvalues $\lambda_k$ of the Dirac operator $D$ are sensitive to breaking of chiral symmetry. In finite volume $V$ the eigenvalue density is the ensemble average
$\rho(\lambda,m) = \sum_{k=1}^{\infty} \langle \delta(\lambda - \lambda_k) \rangle/V$.
The Banks-Casher relation for two flavors $\lim_{\lambda \rightarrow 0} \lim_{m \rightarrow 0} \lim_{V \rightarrow \infty} \rho(\lambda,m) = \Sigma/(2\pi)$, where $\Sigma = - \langle \bar{\psi} \psi \rangle$, connects the eigenvalue density to the fermion condensate. The relation is somewhat formal, in practice the lattice volume must be sufficiently large to allow a large enough number of small eigenvalues to emerge at small mass $m$. Reversing the limits and taking $m \rightarrow 0$ at finite $V$ always gives a zero eigenvalue density, regardless of whether or not the theory has spontaneous symmetry breaking. To test the relation in its simplest form, we measure the eigenvalue density by directly calculating a fixed number of the smallest eigenvalues for each gauge ensemble. As the lattice volume increases, more eigenvalues must be determined to cover an appreciable range. Because of the large numerical cost of eigenvalue determination, we select a subset of gauge configurations well-separated in the Markov chain, to minimize autocorrelation effects. We use the same gauge ensembles as were used to measure the mass spectrum. We show the eigenvalue density for the largest lattice volume $48^3 \times 96$ at the lightest fermion mass $m=0.003$ on the right in Figure~\ref{fig1}. The eigenvalue density increases slowly with $\lambda$ and at this mass the $\rho(\lambda = 0)$ density is quite close to the chiral extrapolation of the directly-measured condensate $\langle \bar{\psi} \psi \rangle$. The quantitative effect of topology and would-be zero modes on the eigenvalue density remains to be investigated, as well as finite-volume effects which are most pertinent at the low end of the eigenvalue spectrum.

An alternate method to extract the fermion condensate from the eigenvalue density is via the mode number. One calculates the eigenvalues $\lambda$ of the Hermitian operator $D^{\dagger} D + m^2$ and determines how many eigenvalues are below some scale $M^2$, namely $\nu(M,m) = V \int_{-\Lambda}^{\Lambda} d \lambda~\rho(\lambda,m)$, where $\Lambda = \sqrt{M^2 - m^2}$. The mode number $\nu(M,m)$ is renormalization-group invariant, a new proof of which was recently given~\cite{Giusti:2008vb}. At leading order one can define an effective fermion condensate $\Sigma_{\rm eff} = (\pi/2V) d\nu(M,m)/d\Lambda$, which in the chiral limit yields the condensate $\Sigma$. It was shown in $N_f=2$ flavor QCD simulations that the mode number has mild dependence on the fermion mass, allowing a straightforward linear extrapolation to the chiral limit at a value of $M$ where finite-volume contamination was not detectable~\cite{Giusti:2008vb}. Separately it was also shown~\cite{Smilga:1993in} that for $N_f=2$ flavors, the 1-loop correction to $\Sigma_{\rm eff}$ in chiral perturbation theory is zero for any choice of scale $\Lambda$, which may explain the weak mass dependence. One possible advantage of this method is that the mode number can be calculated stochastically for a given choice of scale $M$, as opposed to direct exact calculation of the lowest eigenvalues~\cite{Giusti:2008vb}. The number of eigenvalues required to reach a given value of $M$ increases as the lattice volume increases, hence the exact calculation of the mode number becomes ever more numerically costly with the volume. In comparison the numerical cost of the stochastic method may increase more slowly with lattice volume. One advantage of determining the exact eigenvalues is one can {\it post hoc} calculate the mode number for a range of $M$ values, whereas the stochastic approach described in~\cite{Giusti:2008vb} requires the value of $M$ to be set beforehand and is hence less flexible. The mode number was recently used in the context of possible BSM gauge theories to extract the anomalous mass dimension in $SU(2)$ gauge theory with $N_f=2$ Dirac fermions in the adjoint representation~\cite{Patella:2012da} and in $SU(3)$ gauge theory with $N_f=4, 8$ or 12 fermions in the fundamental representation~\cite{Cheng:2013eu}.

\section{Results}

In this first study of the mode number in the sextet gauge theory, we calculated a large number of the small eigenvalues exactly in order to have flexibility in the choice of $M$. Having gained this experience, we plan to extend the study exploiting the stochastic technique. We have a set of gauge ensembles covering fermion masses in the range $m=0.003$ -- $0.008$ and lattice volumes $48^3 \times 96, 32^3 \times 64, 28^3 \times 56$ and $24^3 \times 48$, where the lowest Goldstone mass reached is roughly 0.13 and the smallest Goldstone decay constant is roughly 0.04, in lattice units. All ensembles are at the same lattice spacing corresponding to a bare coupling $\beta = 3.20$, we currently have data at three more couplings for a future more refined analysis. We simulate using staggered fermions, with stout smearing~\cite{Morningstar:2003gk} to reduce taste breaking, the tree-level Symanzik-improved gauge action, and the Rational Hybrid Monte Carlo algorithm~\cite{Clark:2006fx} for $N_f=2$ flavors.

\begin{figure}[thb!]
\begin{center}
\begin{tabular}{cc}
\includegraphics[width=0.48\textwidth]{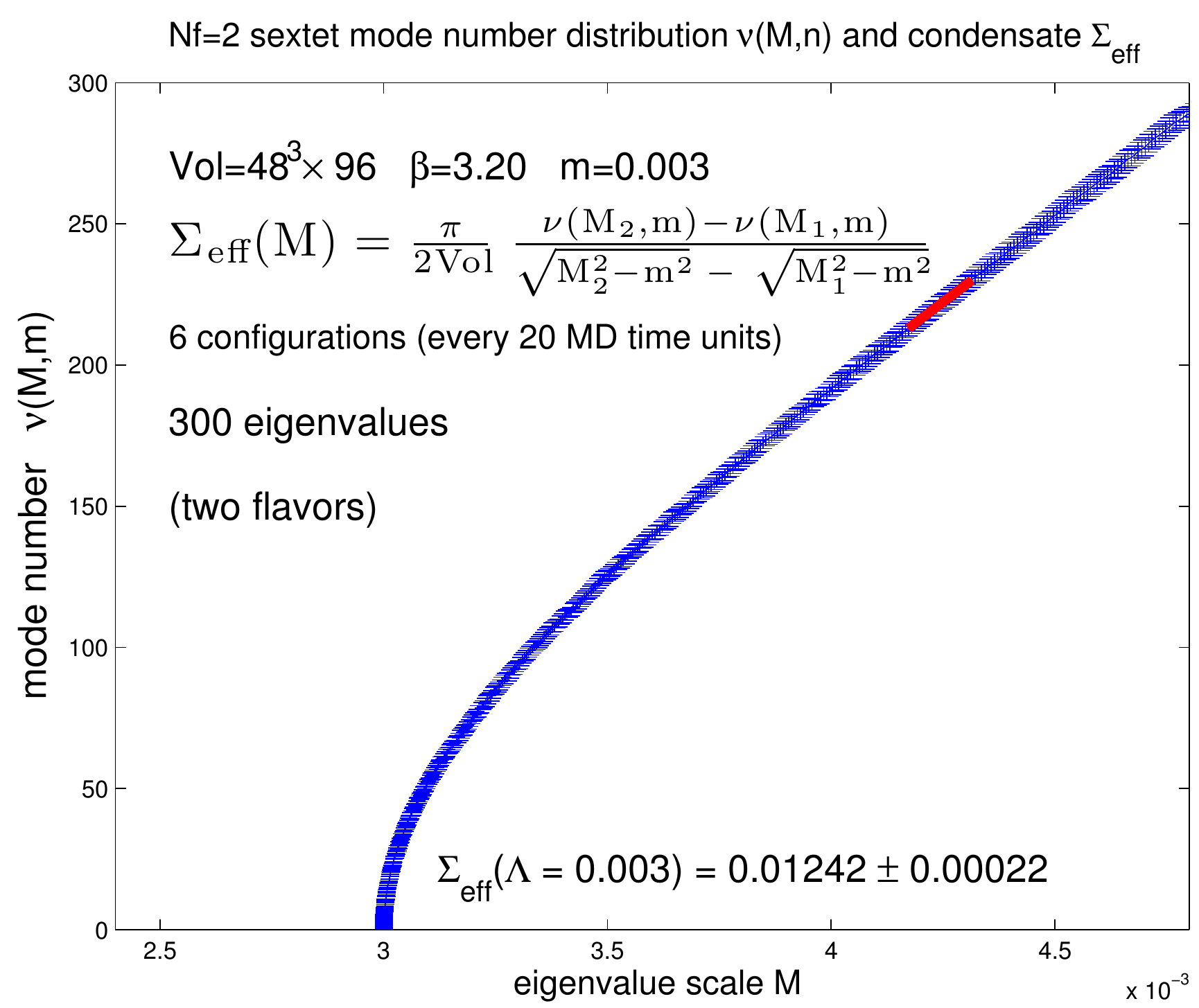}&
\includegraphics[width=0.46\textwidth]{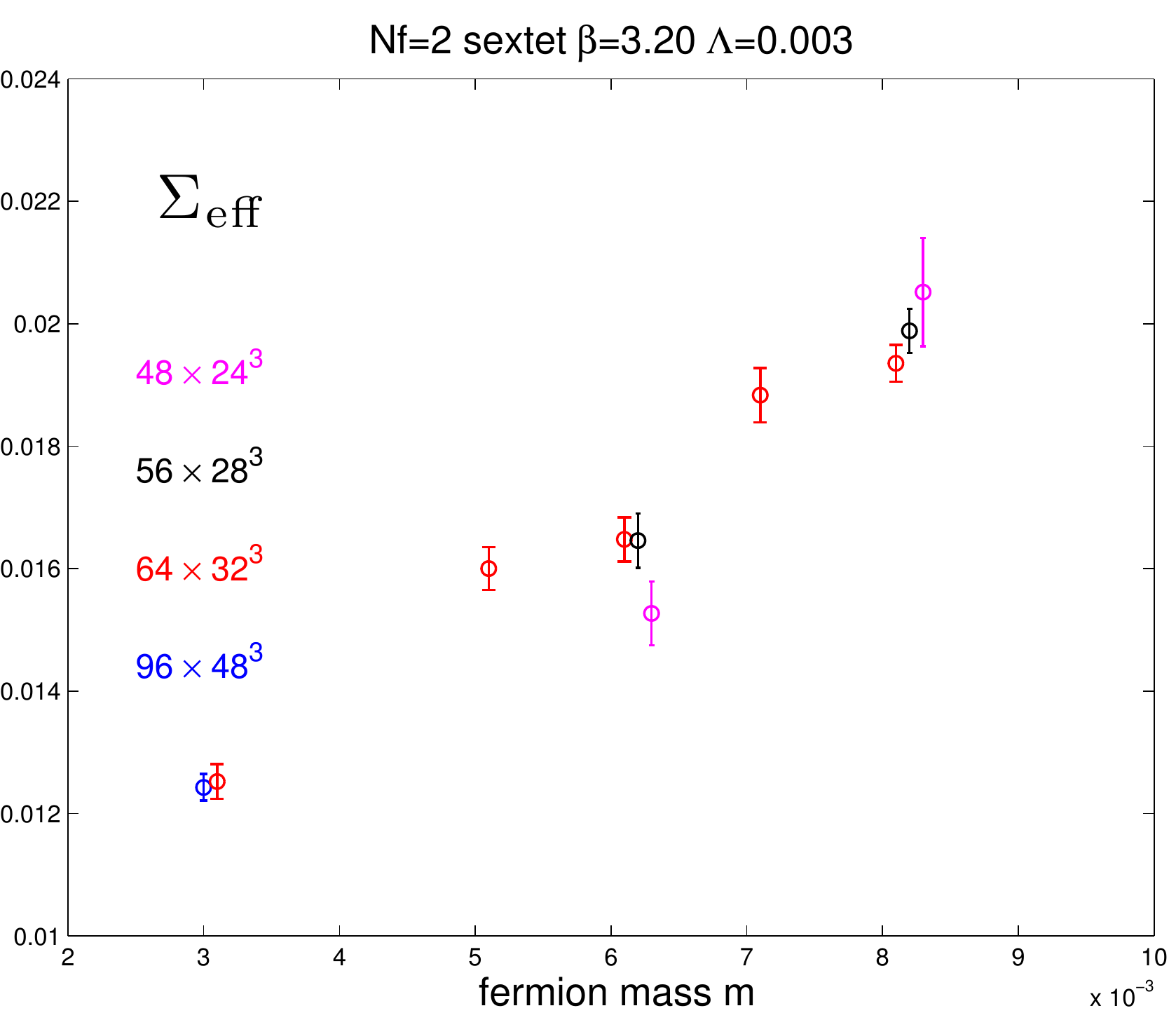}
\end{tabular}
\end{center}
\vskip -0.3in
\caption{\footnotesize (left) The mode number $\nu$ as a function of the eigenvalue scale $M$ as measured on $48^3 \times 96$ lattice volumes. The region where the derivative of the mode number is used to extract the effective condensate $\Sigma_{\rm eff}$ is shown by the red line. The derivative $d \nu/d\Lambda$ is approximated by a finite difference between $M_2$ and $M_1$ centered around $M$. (right) The volume-dependence of the effective condensate $\Sigma_{\rm eff}$, extracted on all ensembles at $\Lambda = 0.003$. The data are slightly offset horizontally for visibility.}
\label{fig2}
\end{figure}

We show on the left in Figure~\ref{fig2} the mode number on a lattice volume $48^3 \times 96$ at the lightest fermion mass. The lowest 300 eigenvalues were determined for each gauge configuration considered, which in this large volume extends out to a maximum eigenvalue scale $M$ around 0.005. The blue band gives the error estimate of the mode number from the jackknife method. Even with relatively few gauge configurations, the mode number is quite accurately measured. We measure on configurations separated by 20 Molecular Dynamics time units to reduce autocorrelation. Motivated by the leading-order linear relationship between $\Sigma_{\rm eff}$ and $\nu$, we define the effective condensate via the derivative $d \nu/d\Lambda$, which we approximate with finite differences around the central value. Deviation from linearity would reflect the increase in $\rho$ moving away from $\lambda = 0$. The choice $\Lambda = 0.003$ is convenient for all ensembles as being in the central eigenvalue region, neither too close to the maximal $M$ value due to the finite number of eigenvalues being calculated, nor too close to the lower end of the eigenvalue spectrum. The location $\Lambda=0.003$ is shown on the left in Figure~\ref{fig2} as the red line where the derivative is calculated.

\begin{figure}[hbt!]
\begin{center}
\begin{tabular}{cc}
\includegraphics[width=0.48\textwidth]{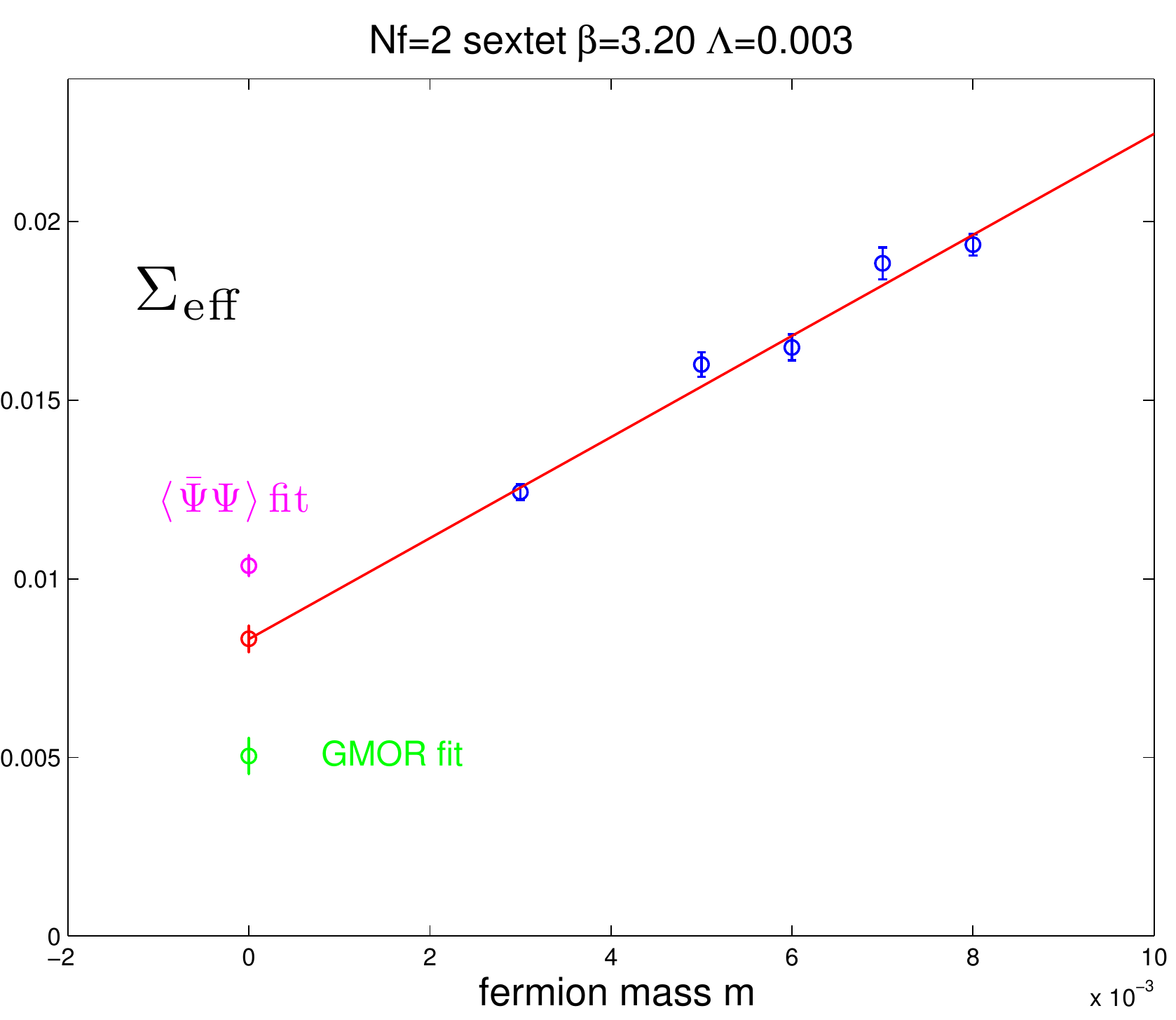}&
\includegraphics[width=0.48\textwidth]{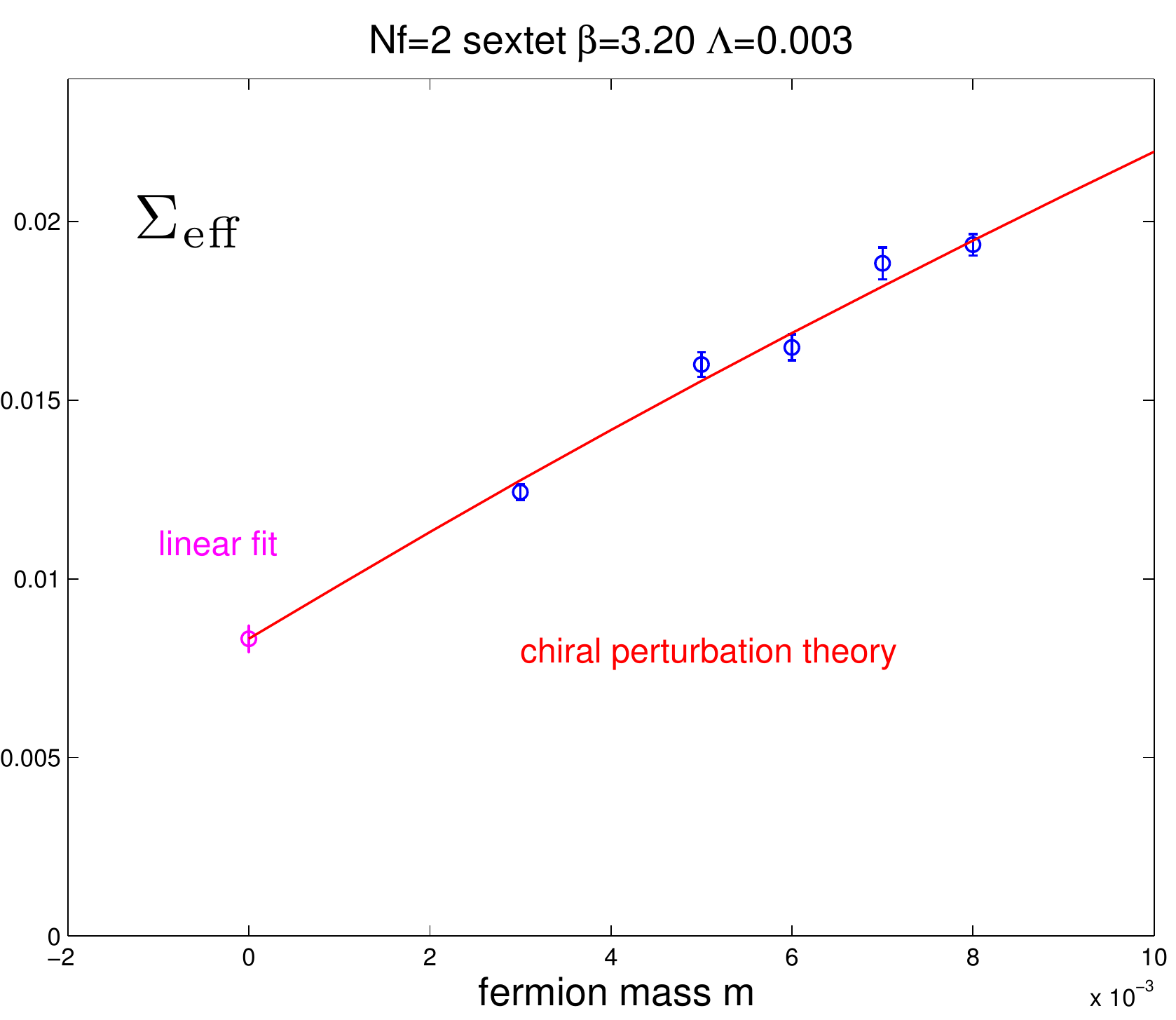}
\end{tabular}
\end{center}
\vskip -0.3in
\caption{\footnotesize (left) A linear extrapolation of $\Sigma_{\rm eff}$ in the fermion mass $m$, using the largest volume determination at each mass. The extrapolation gives $\Sigma = 0.00831(38)$ in the chiral limit with $\chi^2/N_{\rm dof} = 7.1/3$. The fermion condensate values extrapolated from direct measurement of $\langle \bar{\psi} \psi \rangle$ and from the GMOR relation are included for comparison. (right) A comparison of the mass dependence of $\Sigma_{\rm eff}$ with chiral perturbation theory.}
\label{fig3}
\end{figure}

We repeat the analysis for each ensemble at the value $\Lambda = 0.003$, the results are summarized on the right in Figure~\ref{fig2}. At three values of the fermion mass $m$, there is good consistency in the determination of $\Sigma_{\rm eff}$ from different lattice volumes, an empirical indication that the physical volume is large enough to allow a non-zero density of small eigenvalues to emerge. For further analysis, we treat the value of $\Sigma_{\rm eff}$ on the largest lattice volume at each fermion mass as being the infinite-volume result. As shown on the left in Figure~\ref{fig3}, we find the data can be described quite well by linear mass dependence. The extrapolation gives a value for the fermion condensate in the chiral limit which lies between those obtained from the direct measurement of $\langle \bar{\psi} \psi \rangle$ and from the GMOR relation. We can also compare the data with an expansion in the fermion mass from chiral perturbation theory. The analytic result from Osborn et al~\cite{Osborn:1998qb} is
\begin{equation}
\frac{\Sigma_{\rm eff}}{\Sigma} = 1 + \frac{\Sigma}{32 \pi^3 N_F F^4} \left[ 2 N_F^2 |\Lambda| \arctan \frac{|\Lambda|}{m} - 4 \pi |\Lambda| - 
N_F^2 m \log \frac{\Lambda^2 + m^2}{\mu^2} - 4 m \log \frac{|\Lambda|}{\mu} \right]
\label{eq1}
\end{equation}
where the scale is set by $\mu = F^2 \Lambda^2_{\rm mom}/2 \Sigma$ and $\Lambda_{\rm mom}$ is the momentum cutoff (the term $\pi^3$ as above is a correction). In the special case $N_F=2$, there is no $\Lambda$ correction in the limit $m \rightarrow 0$. As we show on the right in Figure~\ref{fig3} the chiral form appears to describe the data quite well, with a roughly linear dependence in this mass range, where the values of $F$ and $\Sigma$ are taken from the simulation results. We do not discuss here the issue of renormalization.

\section{Refinements and outlook}

The numerical cost of exact eigenvalue calculation grows too quickly with the lattice volume to allow significantly increasing the maximum value of the eigenvalue scale $M$ beyond what we show here. Hence the systematic effect of the choice of $\Lambda$ on the determination of the fermion condensate cannot be tested this way. We are currently testing a new generalization of the stochastic method with complete flexibility in the choice of scale.

One possible explanation for the disparity between the various determinations of the chiral condensate in the massless limit is taste breaking inherent at finite lattice spacing. This can be accounted for in staggered chiral perturbation theory as outlined in~\cite{Osborn:2010eq} by inclusion of the leading order taste-breaking operators. Alternatively, running additional simulations at smaller lattice spacing would probe the magnitude of cutoff effects. Another possibility is that the simulations are at too large fermion mass for chiral perturbation theory to be applicable; for example the decay constant $F_{\pi}$ changes significantly over the mass range shown here. In addition, in these simulations the composite scalar is as light as the Goldstone bosons, perhaps requiring chiral perturbation theory to be extended to include the effects of this new light degree of freedom.

The ensembles we use in this analysis are in the $p$-regime, where the system is essentially at infinite volume. An alternative method is to simulate the theory in the $\epsilon$- and $\delta$-regime, where the fermion mass is chosen small enough such that the infinite-volume Goldstone bosons are much lighter than can be accommodated in the available finite volume. One can tune the infinite-volume Goldstone bosons to be lighter than the composite scalar, which is likely to be frozen out, simplifying the analysis. In this regime, the low-lying eigenvalues have a characteristic dependence on the fermion mass and volume which allows the infinite-volume condensate $\Sigma$ to be extracted. Alternatively one can work directly with the $SU(2)$ rotator spectrum. We previously explored the $\epsilon$-regime for $SU(3)$ gauge theory with $N_f=4$ and 8 flavors in the fundamental representation~\cite{Fodor:2009wk}, showing that the eigenvalue distributions supported both theories having spontaneous chiral symmetry breaking. A necessary ingredient in this approach is to calculate separate gauge ensemble averages for each topological sector. We previously studied the topological properties of the sextet theory, such as the index theorem, over a range of lattice spacings in the quenched approximation~\cite{Fodor:2009nh}. The topological behavior in the $p$-regime and its relationship with the low-lying eigenvalues also remains to be fully explored in this model.

\section*{Acknowledgments}
We acknowledge support by the DOE under grant DE-SC0009919, by the NSF under grants 0704171 and 0970137, by the EU Framework Programme 7 grant (FP7/2007-2013)/ERC No 208740, by OTKA under the grant OTKA-NF-104034, and by the Deutsche Forschungsgemeinschaft grant SFB-TR 55. Computational resources were provided by USQCD at Fermilab and JLab, at the UCSD GPU cluster funded by DOE ARRA Award ER40546, by the NSF grant OCI-1053575 at the Extreme Science and Engineering Discovery Environment (XSEDE), and at the University of Wuppertal. Simulations on GPU clusters were facilitated by the CUDA ports of~\cite{Egri:2006zm}. KH wishes to thank the Institute for Theoretical Physics and the Albert Einstein Center for Fundamental Physics at Bern University for their support.

\end{document}